\documentclass[%
 reprint,
superscriptaddress,
nobibnotes,
 amsmath,amssymb,
 aps,
floatfix,
]{revtex4-2}

\usepackage{graphicx}% Include figure files
\usepackage{dcolumn}% Align table columns on decimal point
\usepackage{bm}% bold math
\usepackage{hyperref}% add hypertext capabilities
\usepackage{physics}

\begin{document}

\preprint{APS/123-QED}

\title{Machine Learning Inversion from Small-Angle Scattering for Charged Polymers}

\author{Lijie Ding}
\affiliation{Neutron Scattering Division, Oak Ridge National Laboratory, Oak Ridge, TN 37831, USA}
\author{Chi-Huan Tung}
\affiliation{Neutron Scattering Division, Oak Ridge National Laboratory, Oak Ridge, TN 37831, USA}
\author{Jan-Michael Y. Carrillo}
\affiliation{Center for Nanophase Materials Sciences, Oak Ridge National Laboratory, Oak Ridge, TN 37831, USA}
\author{Wei-Ren Chen}
\affiliation{Neutron Scattering Division, Oak Ridge National Laboratory, Oak Ridge, TN 37831, USA}
\author{Changwoo Do}
\email{doc1@ornl.gov}
\affiliation{Neutron Scattering Division, Oak Ridge National Laboratory, Oak Ridge, TN 37831, USA}
\date{\today}% It is always \today, today,
             %  but any date may be explicitly specified

\begin{abstract}
We develop Monte Carlo simulations for uniformly charged polymers and machine learning algorithm to interpret the intra-polymer structure factor of the charged polymer system, which can be obtained from small-angle scattering experiments. The polymer is modeled as a chain of fixed-length bonds, where the connected bonds are subject to bending energy, and there is also a screened Coulomb potential for charge interaction between all joints. The bending energy is determined by the intrinsic bending stiffness, and the charge interaction depends on the interaction strength and screening length. All three contribute to the stiffness of the polymer chain and lead to longer and larger polymer conformations. The screening length also introduces a second length scale for the polymer besides the bending persistence length. To obtain the inverse mapping from the structure factor to these polymer conformation and energy-related parameters, we generate a large data set of structure factors by running simulations for a wide range of polymer energy parameters. We use principal component analysis to investigate the intra-polymer structure factors and determine the feasibility of the inversion using the nearest neighbor distance. We employ Gaussian process regression to achieve the inverse mapping and extract the characteristic parameters of polymers from the structure factor with low relative error.

\end{abstract}

%\keywords{Suggested keywords}%Use showkeys class option if keyword
                              %display desired
\maketitle

\section{Introduction}
Semiflexible charged polymers \cite{netz2003neutral}, also known as polyelectrolytes \cite{dobrynin2005theory,forster2005polyelectrolytes}, represent an essential class of materials that are fundamental to both biological processes and technological applications\cite{manning1978molecular}. Their unique behaviors emerge from the interplay between molecular flexibility and electrostatic interactions, which are governed by the presence of ionizable groups along their chains. Notable natural examples include DNA \cite{lameh2020controlled,manning1978molecular}, RNA \cite{bloomfield2000nucleic}, and proteins \cite{tanford1962physical}, all of which play pivotal roles in cellular functions. Synthetic polyelectrolytes, on the other hand, have found extensive use in a variety of fields, including water treatment \cite{bolto2007organic}, energy storage \cite{winter2004batteries}, drug delivery \cite{liechty2010polymers}, and responsive materials \cite{stuart2010emerging}. The conformational and dynamic properties of charged polymers are shaped by factors such as charge density, ionic strength of the surrounding environment, and the intrinsic bending stiffness of the polymer chain. A thorough understanding of these properties is crucial for tailoring polyelectrolytes to meet the specific demands of diverse applications. 

To understand the structure and behavior of the charged polymers, both experimental and theoretical approaches have been employed. Experimental techniques such as small-angle scattering\cite{lindner2002neutrons} (SAS) including X-ray scattering\cite{chu2001small} and neutron scattering\cite{chen1986small,shibayama2011small} have proven indispensable for understanding these properties of the charged polymers\cite{nierlich1979small}. Scattering methods provide insights into the nanoscale structure and dynamics of charged polymers, enabling the characterization of key conformational parameters such as radius of gyration, persistence length, and inter- and intra-molecular interactions. Theoretical and computational approaches, including analytical models\cite{netz2003variational,dobrynin1995scaling} and computer simulations, complement experimental efforts by capturing the fundamental physics of charged polymer systems. Techniques such as molecular dynamics\cite{stevens1995nature, gubarev2009scale} (MD) and as Monte Carlo\cite{carlsson2001monte,chodanowski1999monte} (MC) simulations have provided significant insights into polymer configurations, bending rigidity, and electrostatic interactions.

Despite the progress made on both the experimental and theoretical fronts, bridging the scattering function measured in SAS experiments with the polymer parameters used for modeling charged polymers in theory and simulations remains a significant challenge. The difficulties lies in extracting physical quantities about polymer conformation by decoding the scattering function. Recent advances in machine learning (ML) have opened new avenues in scattering analysis, enabling parameter extraction without requiring explicit analytical forms of the scattering function\cite{chang2022machine}. By training ML models on simulation-generated data, it becomes possible to establish an inverse mapping from the scattering function to the underlying model parameters. This approach has shown promise in a variety of systems, including colloids \cite{chang2022machine, tung2023inferring, ding2024machine_correlation}, polymers \cite{tung2022small, ding2024machine, ding2024machine_ladderpolymer}, and lamellar structures \cite{tung2024unveiling, tung2024scattering}. These applications demonstrate the potential of ML to bridge the gap between experimental scattering data and theoretical models, providing a robust framework for parameter extraction in complex systems.

In this work, we introduce such inversion by ML approach for the charged polymer system, where the data are generated using MC simulations. The polymer configuration is governed by the intrinsic bending stiffness, charge density and salt concentration of surrounding medium. We first investigate the effects of these key variables on polymer conformation and then calculate the intra-polymer structure factor. To assess the feasibility of inversion, we perform principal component analysis on the scattering data and quantify the such feasibility using nearest neighbor distance of the polymer parameters in the structure factor space. Finally, we employ Gaussian process regression (GPR) to extract both conformational and energy-related parameters of the polymers from the structure factor, demonstrating the accuracy and robustness of this approach.

\section{Method}
\subsection{Charged polymer in ironic fluid}
We model the polymer as a chain of $N$ connected bonds with fixed length $l_b$, such that the joint connecting bonds $i-1$ and $i$ is $\mathbf{r}_i$ and the tangent of bond $i$ is $\mathbf{t}_i \equiv (\mathbf{r}_{i+1} - \mathbf{r}_i) / l_b$. The polymer energy is given by:
\begin{equation}
    E = \sum_{i=0}^{N-2} \frac{\kappa}{2}\frac{(\mathbf{t}_{i+1} - \mathbf{t}_i)^2}{l_b} - \sum_{i=0}^{N-1}\sum_{j \neq i}^{N-1}\frac{A}{r_{ij}}e^{-r_{ij}/\lambda_D}
    \label{equ:energy}
\end{equation}
where $\kappa$ is the bending modulus, $\frac{A}{r_{ij}}e^{-r_{ij}/\lambda_D}$ is the Yukawa potential, or screened Coulomb potential\cite{hansen2013theory,gubarev2009scale}, that models the charge interaction, $A$ is the interaction strength between charged monomers, $\lambda_D$ is the Debye screening length\cite{israelachvili2011intermolecular}, and $r_{ij} = |\va{r}_i - \va{r}_j|$ is the distance between joints $i$ and $j$. In addition, the self-avoidance of the polymer is enforced by adding hard sphere interaction of diameter $l_b$ between different joints. The interaction strength $A=\frac{(\sigma_e l_b)^2}{4\pi \epsilon}$ is directly related to the charge density of the polymer $\sigma_e$, where $\epsilon$ is the dielectric constant of the medium. The the Debye screen length $\lambda_D = \sqrt{\frac{\epsilon k_B T}{2 e^2 I}}$, where $k_B$ is the Boltzmann constant, $T$ is the system temperature, $e$ is elementary charge, $I=\frac{1}{2}\sqrt{z_i^2 n_i}$ is the ionic strength, in which $z_i$ and $n_i$ are charge number of the number density of ion species $i$, respectively.

\subsection{Monte Carlo simulation}
To calculate the conformational properties of the charged polymer at equilibrium, we sample the configuration space of the charged polymers using the off-lattice Markov Chain Monte Carlo (MCMC) method\cite{ding2024off} we previously developed, this off-lattice method provide accurate calculation of the polymer conformation and overcome the orientational bias rooted in the lattice model\cite{tung2024discretized}. The polymer configuration $\left\{\va{r}_0, \va{r}_1 \dots \va{r}_{N-1} \right\}$ is updated using two MC moves: continuous crankshaft and pivot. Crankshaft picks two random joints on the polymer chain and rotate all the bonds between them for a random angle. Pivot randomly select one joint on the chain and rotate the preceding sub-chain within a cone centering at the original orientation. These two moves combined allows full exploration of the polymer configuration with the contour length fixed and the polymer conformation calculated using this algorithm has been bench marked against theoretical calculations.

To better characterize and understand the conformation of the charged polymer, we calculate the radius of gyration, bond angle correlation and structure factor of the polymer. The radius of gyration square is $R_g^2 = \frac{1}{2}\left<r_{ij}^2\right>_{ij}$, the $\left<\dots \right>_{ij}$ denotes the average of all pair of joints. The bond-bond correlation is $\left<\cos(\theta(s))\right> = \left<\vu{t}_i\cdot \vu{t}_{i+s}\right>_i$ where $\left<\dots \right>_i$ denotes the average over all bonds and s represents the contour distance between two bonds along the polymer chain. Finally, the isotropic intra-polymer structure factor\cite{lindner2002neutrons,chen1986small} is given by:
\begin{equation}
    S(q) = 1+\frac{1}{N^2}\sum_{i=0}^{N-1}\sum_{j\neq i}^{N-1} \frac{\sin(q r_{ij})}{q r_{ij}}
    \label{equ:Sq}
\end{equation}
where $q$ is the magnitude of the scattering vector. When running the MCMC simulation, we first randomize the system by running 2000 MC sweeps at inverse temperature $\beta=1/k_B T = 0$, then tempering the system for another 2000 MC sweeps while gradually decrease the temperature to $\beta=1$. We sample the polymer configuration and calculate the average of the conformation parameters for while running for another 4000 MC sweeps, each MC sweep is consist of $N$ crankshafts and $N$ pivot updates. We use natural unit in our simulation where energy is in unit of $k_B T=1$ and length is in unit of $l_b=1$ such that the polymer contour length $L=N l_b = N$. We use degree of discretization $L=500$ for all of our simulations.

\subsection{Principle component analysis}
\label{ssec: PCA}
To study the relationship between structure factor $S(q)$ and the polymer parameters including radius of gyration $R_g^2$, end-to-end distance $R^2$, bending stiffness $\kappa$ and interaction strength $A$ for various screening distance $\lambda_D$, we generate a data set consisting of $4,000$ combinations of $(\kappa, A, \lambda_D)$ and corresponding $\log {S(q)}$ and carry out the principal component analysis for the data sets. The $S(q)$ is calculated for $100$ $q\in [10^{-1},1]$, uniformly placed in log scale, and $\kappa \sim U(5,50)$, $A\sim U(0,10)$ and $\lambda_D \sim U_d(1,10)$, where $U(a,b)$ is the uniform distribution in the interval $[a,b]$ and $U_d(a,b)$ is the discrete uniform distribution.
Similar to previous work\cite{chang2022machine}, we use singular value decomposition (SVD) to find the three most important basis of the $4,000 \times 100$ matrix $\vb{F}  = \{\log{S(q)}\}$, such that $\vb{F} = \vb{U}\vb{\Sigma} \vb{V}^T$. The diagonal entries of $\Sigma^2$ are proportional to the weight of the variance of the projection of $\vb{F}$ onto each principal vectors of $\vb{V}$. Projecting the $\vb{F}$ to the first few basis provide a way to analyze the $\vb{F}$ is a dimension reduced space. A useful tool to study the distribution of the polymer parameters $\vb{Y} = \{(\kappa, A, R_g^2/L^2, R^2/L^2)\}$ is to calculate the nearest-neighbor distance of $\zeta \in \{\kappa, A, R_g^2/L^2, R^2/L^2\}$ on the $\vb{F}$ manifold. For n-number of vectors, ${\vb{x}_1,\vb{x}_2,\dots, \vb{x}_n}$, the first nearest neighbor is defined as $NN_1(\vb{x}_i) = \arg\!\min_{\vb{x}_j\neq \vb{x}_i} |\vb{x}_j - \vb{x}_i|$, similarly, the second nearest neighbor is $NN_1(\vb{x}_i) = \arg\!\min_{\vb{x}_j\neq \vb{x}_i, NN_1(\vb{x}_i)} |\vb{x}_j - \vb{x}_i|$, we define the normalized nearest neighbor distance $D_{NN}$ for the $\zeta(\vb{x})$ as:
\begin{equation}
    D_{NN}(\zeta) = \frac{\left<2\zeta(\vb{x})-\zeta(NN_1(\vb{x}))-\zeta(NN_2(\vb{x}))\right>_{\vb{x}}}{(\max_{\vb{x}}(\zeta) - \min_{\vb{x}}(\zeta))/2}
\end{equation}
where $\left<\dots\right>_{\vb{x}}$ is the average over all $\vb{x}$.

\subsection{Gaussian process regression}
To perform the inverse mapping from the scattering function, $\vb{x} = \log{S(q)}$, to the system parameters, or inversion targets $\vb{y} = (\kappa, A, R_g/L^2, R^2/L^2)$, we employ a Gaussian Process Regression (GPR) model trained on data generated through Monte Carlo (MC) simulations. Under the framework of GPR \cite{williams2006gaussian, wang2023intuitive}, the goal is to obtain the posterior distribution $p(\vb{Y}_*|\vb{X}_*,\vb{X},\vb{Y})$ for the function output $\vb{y}$. In this setup, the training and test sets are defined as $\vb{X} = \{\log S(q)\}_{train}$ and $\vb{X}_* = {\log S(q)}_{test}$, respectively, while $\vb{Y}$ and $\vb{Y}*$ correspond to the inversion targets $(\kappa, A, R_g/L^2, R^2/L^2)$. GPR assumes a Gaussian process prior over the regression function, $g(\vb{x}) \sim GP(m(\vb{x}), k(\vb{x}, \vb{x}'))$, where $m(\vb{x})$ is the prior mean function, and $k(\vb{x}, \vb{x}')$ is the covariance kernel. The joint distribution for the Gaussian process is expressed as follows:

\begin{equation} \mqty(\vb{Y} \\ \vb{Y}_*) \sim \mathcal{N}\left( \mqty[m(\vb{X}) \\ m(\vb{X}_*)], \mqty[k(\vb{X},\vb{X}) & k(\vb{X},\vb{X}_*) \\ k(\vb{X}_*,\vb{X}) & k(\vb{X}_*,\vb{X}_*)] \right) 
\label{equ: Gaussian_process} 
\end{equation}

Here, we use a constant prior mean function $m(\vb{x})$, while the kernel function is modeled as a combination of a Radial Basis Function (RBF) and a white noise term:
\begin{equation} 
k(\vb{x}, \vb{x}') = \exp\left(-\frac{|\vb{x} - \vb{x}'|^2}{2l}\right) + \sigma \delta(\vb{x}, \vb{x}'), 
\end{equation} 
where $l$ represents the correlation length, $\sigma$ is the variance of the observational noise, and $\delta$ is the Kronecker delta function. These hyperparameters are optimized during training using the simulation data. In practice, we utilize the scikit-learn\cite{scikit-learn, sklearn_api} Gaussian Process library due to its convenience and efficiency.

\section{Results}
We first study the affect of each polymer parameters on the conformation of the polymer, then investigate the scattering function of the charged polymer, where we also shows the principal component analysis of the our data set $\vb{F}=\{\log S(q)\}$, we then discuss the feasibility of inversion based on the SVD of $\vb{F}$. With the feasibility established, we finally test of our trained GPR for the inversion.

\subsection{Variation of polymer conformation}
\begin{figure}
    \centering
    \includegraphics[width=\linewidth]{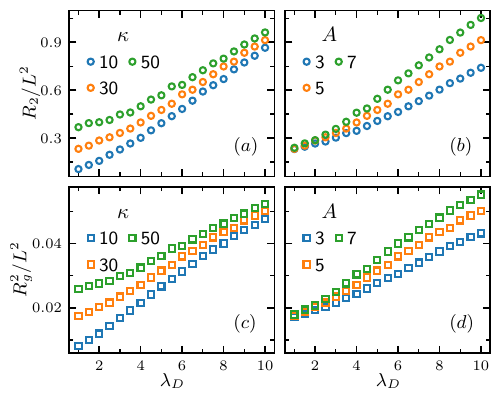}
    \caption{Radius of gyration $R_g^2$ and end-to-end distance $R^2$ of the charged polymer versus various bending stiffness $\kappa$, charge interaction strength $A$ and screen length $\lambda_D$. (a) Normalized end-to-end distance $R^2/L^2$ versus screen length $\lambda_D$ for various bending stiffness $\kappa$. (b) $R^2/L^2$ versus screen length $\lambda_D$ for various charge interaction strength $A$. (c) and (d), similar to (a) and (b), respectively, but for normalized radius of gyration $R_g^2/L^2$} 
    \label{fig:R2_Rg2}
\end{figure}

Both the local bond-to-bond bending and long-range charge interaction contribute to the stiffness of the entire polymer. Such stiffness will affect the overall size of the charged polymer, which can be captured by the radius of gyration $R_g^2$ and end-to-end distance $R^2$. Fig.~\ref{fig:R2_Rg2}(a) and (c) shows both the $R_g^2$ and $R^2$ increases with screening length $\lambda_D$ and bending stiffness $\kappa$, and intuitively, the affect of $\kappa$ on both $R_g^2$ and $R^2$ are more significant when $\lambda_D$ is small, as the $R_g^2$ and $R^2$ versus $\lambda_D$ curves for different $\kappa$ start to converge as the $\lambda_D$ increases. On the contrary, while $R_g^2$ and $R^2$ also increases with larger charge interaction strength $A$, these curves diverges as $\lambda_D$ increase, that is because the increasing screening length $\lambda_D$ amplifies the affect of charge interaction.

When the polymer is only subjected to the bending $\kappa$, or in the case of $A=0$. The polymer is a classic semiflexible polymer whose bond angle correlation can be described by a single exponential decay:
\begin{equation}
    \left<\cos\theta(s)\right> = e^{-s/\lambda_0}
    \label{equ:lam0}
\end{equation}
where $\lambda_0$ is the persistent length. $s$ is the bond-bond distance along the polymer contour. However, as pointed out in previous study\cite{gubarev2009scale}, the charge interaction introduces new length scales, as a result, the bond angle correlation can be described by:
\begin{equation}
    \left<\cos\theta(s)\right> = (1-\alpha)e^{-s/\lambda_1} + \alpha e^{-s/\lambda_2}
    \label{equ:lam1_lam2}
\end{equation}
$\lambda_1$ and $\lambda_2$ corresponds to two different length scale, it is also notable that the effective bending rigidity can be calculated by $\lambda_e = \lambda_2/\alpha$\cite{gubarev2009scale}. 

\begin{figure}
    \centering
    \includegraphics[width=\linewidth]{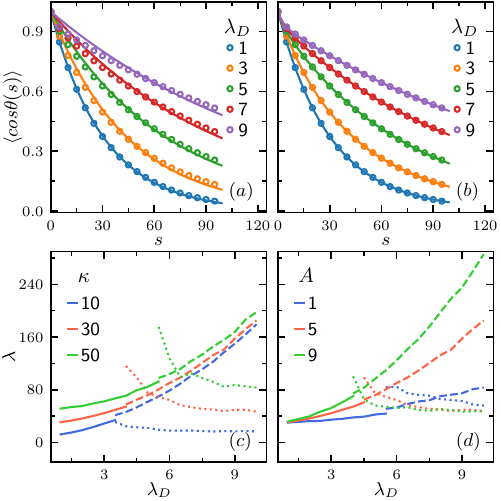}
    \caption{Different length scales of the charged polymer, fitted using both single length scale and double length scale model. (a) Bond angle correlation $\left<\cos\theta(s)\right>$ for various screening length $\lambda_D$ with $\kappa=30$, $A=5$, solid lines are fitted using single length scale Equ.~\eqref{equ:lam0}. (b) similarly, but fitted using double length scale Equ.~\eqref{equ:lam1_lam2}. (c) Three persistent length $\lambda_0$ for solid line, $\lambda_1$ for dashed line and $\lambda_e$ for dotted line, versus screening length $\lambda_D$ for various $\kappa$ with $A=5$. (d) Similar to (c), but for various $A$ with $\kappa=30$.}
    \label{fig:two_length_scale}
\end{figure}

Fig.~\ref{fig:two_length_scale}(a) shows the bond angle correlation function $\left<\cos\theta(s)\right>$ for various screening length $\lambda_D$, and the fitted line are calculated using the single scale model as in Equ.~\eqref{equ:lam0}. As the $\lambda_D$ increases, the single scale model fitting start to diverge from the data point, indicating the necessity of switching to the double length scale model Equ.~\eqref{equ:lam1_lam2}, Fig.~\ref{fig:two_length_scale}(b) shows such fitting results, and the two length scale model can still describe the decay of $\left<\cos\theta(s)\right>$ at large $\lambda_D$. 

Fig.~\ref{fig:two_length_scale} (c) show all three length scales $\lambda_0$, $\lambda_1$ and $\lambda_e$ versus screening length $\lambda_D$ for various bending stiffness $\kappa$. At low $\lambda_D$ the one length scale still fit the bond angle correlation data, and increases with increasing $\lambda_D$. When switching to two length scale model, the long length scale $\lambda_1$ increases with increasing $\lambda_D$, while the short length scale $\lambda_e$ decreases and deviate from $\lambda_1$ then plateaus. The plateau value increases with bending stiffness $\kappa$. Fig.~\ref{fig:two_length_scale} (d) shows similar result but for various charge interaction strength $A$. Similar to its affect on the end-to-end distance and radius of gyration, $A$ amplify the effect of increasing $\lambda_D$, while the short length scale $\lambda_e$ plateaus at similar value for various $A$, confirming it is corresponding to the bending stiffness $\kappa$.

\subsection{Scattering factor of the polymers}
\begin{figure}
    \centering
    \includegraphics[width=\linewidth]{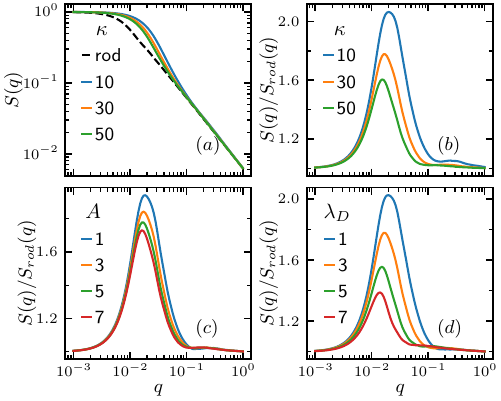}
    \caption{Variation of the structure factor of the charged polymer. (a) Structure factor $S(q)$ for various bending stiffness $\kappa$ with $\lambda_D=3$, $A=5$ and rod effectively representing the $\kappa=\infty$ case. (b) Structure factor $S(q)$ normalized by the rod's structure factor $S_{rod}(q)$ for various $\kappa$. (c) $S(q)/S_{rod}(q)$ for various charge interaction strength $A$ with $\kappa=30$, $\lambda_D=3$. (d) $S(q)/S_{rod}(q)$ for various screening length $\lambda_D$ with $\kappa=30$, $A=5$.}
    \label{fig:Sq}
\end{figure}

We then turn to the inter-polymer structure factor. As a comparison, we also calculate the structure factor of a solid rod, whose polymer configuration is $\va{r}_i = i\vu{x}$, with all bonds pointing to the same direction. Fig.~\ref{fig:Sq}(a) shows the variation of structure factor $S(q)$ for various bending stiffness $\kappa$. Comparing to the solid rod, the polymer structure factor shows a bump at a structure vector $q$ range comparable to its radius of gyration. Fig.~\ref{fig:Sq} (b) shows the structure factor of the polymer divided by the rod $S(q)/S_{rod}(q)$, where the bump is better shown. As the the bending stiffness $\kappa$ increases, the peak in $S(q)/S_{rod}(q)$ lowers and the corresponding $q$ value also decrease, indicating a increases of the characteristic length. Fig.~\ref{fig:Sq}(c) and (d) shows the $S(q)/S_{rod}(q)$ for various charge interaction strength $A$ and screening length $\lambda_D$, both shows similar effects on the structure factor of the polymer as they make the polymer more extended and stiff.

\begin{figure}
    \centering
    \includegraphics[width=\linewidth]{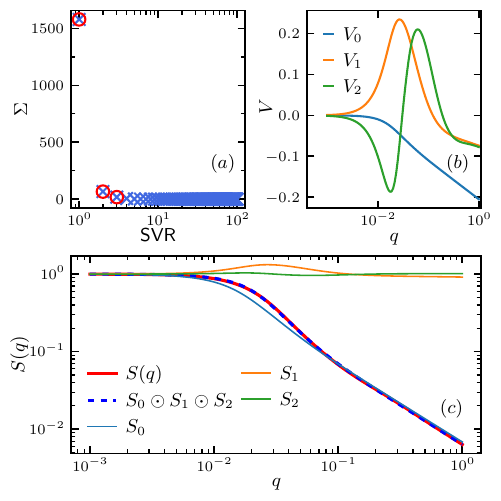}
    \caption{Singular Value Decomposition of the structure factor data set $\vb{F}=\{\log{S(q)}\}$. (a) Singular value $\Sigma$ versus Singular Value Rank (SVR), with top 3 rank highlighted in red circle. (b) First 3 singular vectors $V_0$,$ V_1$ and $V_2$. (c) Decomposition of the $\log{S(q)}$ with $\kappa=10$, $A=5$, $\lambda_D=3$, $\log(S_0),\log(S_1)$ and $\log(S_2)$ are projection of $\log{S(q)}$ on to the $V_0$, $V_1$ and $V_2$, respectively.}
    \label{fig:SVD_basis_Sq}
\end{figure}

To better analyze the structure factor of the charged polymer, we carry out principle component analysis described in Sec.~\ref{ssec: PCA}. By decomposing the $\vb{F}=\{\log{S(q)}\}$ in to $\vb{F} = \vb{U}\vb{\Sigma} \vb{V}^T$. We find that the singular value $\Sigma$ decays rapidly versus its rank, as shown in Fig.~\ref{fig:SVD_basis_Sq}(a), indicating we can represent the $\log{S(q)}\in\vb{F}$ using few basis. Fig.~\ref{fig:SVD_basis_Sq}(b) shows the first 3 singular vectors, and Fig.~\ref{fig:SVD_basis_Sq}(c) shows the projection of a structure factor $S(q)$ on to each basis, and the reconstruction from only the 3 basis closely match the original $S(q)$. This decomposition will allow us to further determine the feasibility of extracting these polymer parameters from the structure factor.

\subsection{Feasibility for Machine Learning inversion}
\begin{figure}
    \centering
    \includegraphics[width=\linewidth]{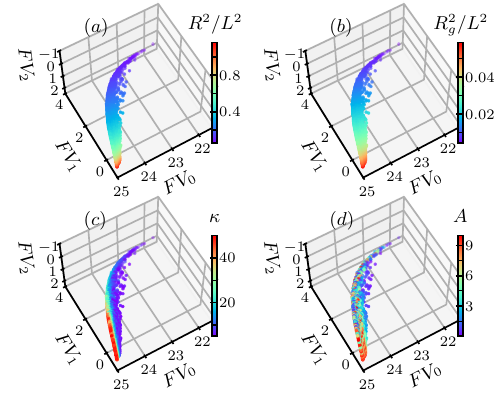}
    \caption{Distribution of the polymer parameters $(R^2/L^2, R_g^2/L^2, \kappa, A)$ in the SVD space spanned by $(V_0, V_1, V_2)$. (a) End-to-end distance divided by length square $R^2/L^2$, (b) Radius of gyration square divided by length square $R_g^2/L^2$. (c) Bending stiffness $\kappa$. (d) Charge interaction strength $A$.}
    \label{fig:SVD_feature}
\end{figure}

While it is straight forward to calculate the structure factor $S(q)$ from the polymer parameters including, length $L$, bending stiffness $\kappa$, charge interaction strength $A$ and screening length $\lambda_D$, and calculate the end-to-end distance $R^2$ and radius of gyration $R_g^2$ using MC simulation. The feasibility of doing the inversion is to be further assessed. Fig.~\ref{fig:SVD_feature} shows the distribution of $(R^2/L^2, R_g^2/L^2, \kappa, A)$ in the structure factor space. This mapping is achieved by projecting all of the structure factor $\log{S(q)}\in\vb{F}$ into the space spanned by the first 3 singular vectors $(V_0,V_1,V_2)$, the corresponding 3 coefficient of each $\log{S(q)}$ corresponds to a single point in the $\mathcal{R}^3$ space. As shown in Fig.~\ref{fig:SVD_feature}(a-c), the end-to-end distance $R^2/L^2$, radius of gyration $R_g^2/L^2$ and bending stiffness $\kappa$ are all well spread out on in $FV$ manifold, indicating they are eligible to be extracted from the structure factor. Fig.~\ref{fig:SVD_feature}(d) shows the distribution of charge interaction strength $A$ and it is unclear if it can be extracted due to some randomness in the distribution.

\begin{figure}[!h]
    \centering
    \includegraphics[width=\linewidth]{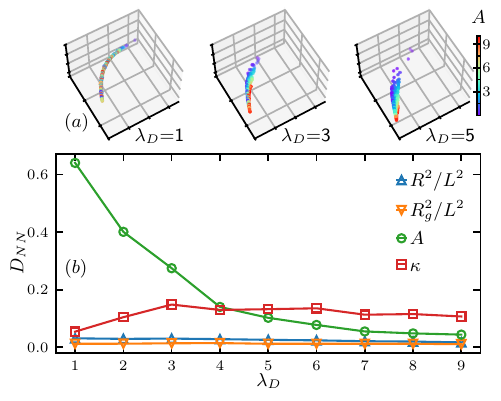}
    \caption{Nearest neighbor distance analysis of the charge interaction strength $A$. (a) Value distribution of $A$ in the SVD space for various slices of screening length $\lambda_D$, the axis are the same as in Fig.~\ref{fig:SVD_feature}. (b) Nearest neighbor distance $D_NN$ for various polymer parameters versus different slices of the data $\vb{F}$ separated by the $\lambda_D$ value.}
    \label{fig:SVD_NND}
\end{figure}

Intuitively, when then screening length $\lambda_D$ is very small, the effect of the charge interaction become negligible, prevent from from showing meaningful impact on the structure factor $S(q)$, thus it is not expected to have the $A$ feasible for extraction from the $S(q)$ at low $\lambda_D$. To quantify this feasibility, we slice the structure factor data set $\vb{F}=\{\log{S(q)}\}$ into different slices for different screening length $\lambda_D$, and calculate the nearest neighbor distance for each slice. As shown in Fig.~\ref{fig:SVD_NND}(a), we plot 3 slices of the charge interaction strength $A$ distribution, and the randomness reduces as the screening length $\lambda_D$ increases. Quantitatively, the Fig.~\ref{fig:SVD_NND}(b) shows the nearest neighbor distance $D_{NN}$ for each polymer parameters and the $D_{NN}(A)$ is much larger than the others when the screening length $\lambda_D$ is small, then it decays to lower value as the $\lambda_D$ increases, lead to more significant impact of the charge interaction strength $A$ on the polymer conformation. This indicate the charge interaction strength $A$, which directly related to the charge density of the polymer, is still extractable if the screening length is large enough.

\subsection{Extraction of the polymer parameters}
\begin{figure}
    \centering
    \includegraphics[width=\linewidth]{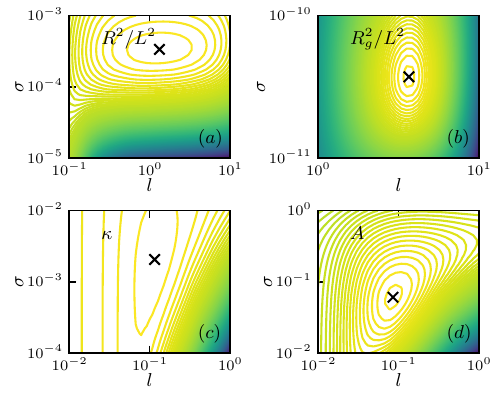}
    \caption{Log marginal likelihood contour of hyperparameters correlation length $l$ and noise level $\sigma$ for various polymer parameters, with optimized value marked with black cross. (a) End-to-end distance $R^2/L^2$. (b) Radius of gyration $R^2/L^2$. (c) Bending stiffness $\kappa$. (d) Charge interaction strength $A$.}
    \label{fig:LML}
\end{figure}

With the feasibility for inversion and corresponding conditions established for the polymer parameter $(R^2/L^2, R_g^2/L^2, \kappa, A)$, we train the GPR using $70\%$ of the entire data set $\vb{F}=\{\log{S(q}\}$as training set $\{\log{S(q}\}_{train}$, and then test the trained GPR using the rest $30\%$ data $\{\log{S(q}\}_{test}$ by comparing the actual polymer parameters with the ones extracted from the structure factor $S(q)$. To obtain the trained regressor, we need to find the optimized hyperparameters $(l,\sigma)$ for each inversion target, or polymer parameters. We search for the $(l,\sigma)$ that maximize the log marginal likelihood\cite{williams2006gaussian}, which are shown in Fig.~\ref{fig:LML}.

\begin{figure}
    \centering
    \includegraphics[width=\linewidth]{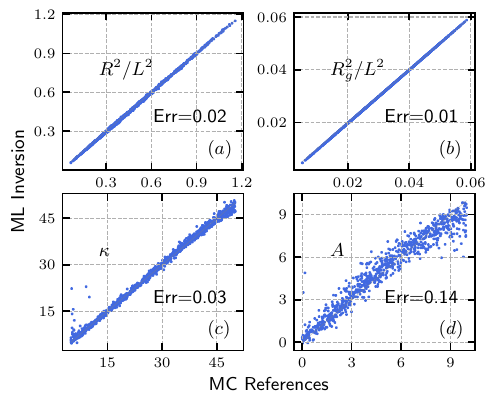}
    \caption{Comparison between polymer parameter extracted from structure factor and input or direct calculation from MC simulation. (a) End-to-end distance $R^2/L^2$. (b) Radius of gyration $R^2/L^2$. (c) Bending stiffness $\kappa$. (d) Charge interaction strength $A$. (a), (b) and (c) utilized all range of $\vb{F}$ and (d) only used data with $\lambda_D\geq 4$.}
    \label{fig:GPR_prediction}
\end{figure}

Fig.~\ref{fig:GPR_prediction} shows the comparison between polymer parameters $((R^2/L^2, R_g^2/L^2, \kappa, A))$ obtained from ML inversion and the corresponding reference used in or calculated through MC simulation. We note that due to the high nearest neighbor distance $D_{NN}(A)$ of charge interaction strength at low screening length $\lambda_D$, we only used data with $\lambda_D\geq4$ for the inversion of $A$. Nevertheless, the data agree well, and lie closely alone the diagonal line, with relatively low error, which for polymer parameter $\zeta$, the relative error between MC reference $\zeta_{MC}$ and ML inversion $\zeta_{ML}$ is estimated by $Err = \left<|\zeta_{MC}-\zeta_{ML}|/max(\zeta_{MC},\zeta_{ML}) \right>$, where $\left<\dots\right>$ here is average over all data points. The relative error are annotated on each panel of the Fig.~\ref{fig:GPR_prediction} and shows very high precision for $((R^2/L^2, R_g^2/L^2, \kappa)$ and good precision for $A$.

\section{Conclusions}
In this work, we apply the off-lattice MC simulation for semiflexible polymer to study the charged polymers, and investigate the ML inversion from scattering for such polymer. We model the polymer using a chain of connected bonds, and the polymer energy is consist of both bending energy and screened Coulomb interaction, which are proportional to the bending stiffness $\kappa$ and charge interaction strength $A$, respectively. The charge interaction range is determined by the screen length $\lambda_D$. We first study the polymer conformation, where the polymer size, quantified by the end-to-end distance $R^2$ and radius of gyration $R_g^2$, increases with $\kappa$, $A$ and $\lambda_D$. The bond angle correlation function transit from single length scale to double length scale as the screening length $\lambda_D$ increases. We calculate the intra-polymer structure factor $S(q)$ of the charged polymer, compare it to which of the solid rod, and shows the $S(q)$ is sensitive to all three polymer parameters $\kappa$, $A$ and $\lambda_D$. We calculate the $S(q)$ for a wide range of $\kappa$, $A$ and $\lambda_D$, then carry out principal component analysis using singular value decomposition to find the singular vectors, which allow us to do dimension reduction of the structure factor. In addition, we investigate the feasibility for inversion from scattering for both the conformation parameters: end-to-end distance $R^2$ and radius of gyration $R_g^2$, and the energy parameters: bending stiffness $\kappa$ and charge interaction strength $A$. We quantify the feasibility using nearest neighbor distance $D_{NN}$, and found the $R^2$, $R_g^2$ and $\kappa$ are eligible for wide range of screening length $\lambda_D$ and the charge interaction strength $A$ is eligible for inversion from structure factor when the $\lambda_D$ is large enough. Finally, we use GPR to obtain the inverse mapping from structure factor $S(q)$ to polymer parameters $(R^2,R_g^2,\kappa,A)$ by optimizing the hyperparameters using a training data set, apply the inversion GPR to extract polymer parameters from structure factor for a test data set, and compare the ML extracted value to the MC reference, they agree well, and low relative errors are achieved.

Our approach provides an unique method to obtain the bending stiffness and the charge density $\sigma_e$, which is directly related to the charge interaction strength $A=\frac{(\sigma_e l_b)^2}{4\pi \epsilon}$ using the scattering data. A natural next step would be to carry out SANS experiment for some charged polymer sample, and apply our approach on the experimentally measured SANS data. In addition, this framework can be expanded to the study of more complicated charged polymer systems including charge-patterned polypeptide\cite{dinic2024effects}, alternating copolymers\cite{yi2019alternating} and zwitterionic patterned polymers\cite{zheng2017applications}. To study these system, it is required to model the polymer energy accordingly. It is natural to introduce variable charge interaction strength $A$ for different monomer segments to model the charge pattern and polarity, and a screened dipole-diple interaction can be used for modeling the zwitterionic polymer.

\begin{acknowledgments}
This research was performed at the Spallation Neutron Source and the Center for Nanophase Materials Sciences, which are DOE Office of Science User Facilities operated by Oak Ridge National Laboratory. This research was sponsored by the Laboratory Directed Research and Development Program of Oak Ridge National Laboratory, managed by UT-Battelle, LLC, for the U. S. Department of Energy. The ML aspects were supported by by the U.S. Department of Energy Office of Science, Office of Basic Energy Sciences Data, Artificial Intelligence and Machine Learning at DOE Scientific User Facilities Program under Award Number 34532. Monte Carlo simulations and computations used resources of the Oak Ridge Leadership Computing Facility, which is supported by the DOE Office of Science under Contract DE-AC05-00OR22725.
\end{acknowledgments}

% The \nocite command causes all entries in a bibliography to be printed out
% whether or not they are actually referenced in the text. This is appropriate
% for the sample file to show the different styles of references, but authors
% most likely will not want to use it.
%\nocite{*}

\bibliography{apssamp}% Produces the bibliography via BibTeX.

\end{document}